\documentclass[12pt, preprint]{aastex}

\newcommand{\lunits}{$\rm erg~s^{-1}$}
\newcommand{\funits}{$\rm erg~cm^{-2}~s^{-1}$}
\newcommand{\cunits}{$\rm cm^{-2}$}
\newcommand{\chandra}{{\it Chandra~}}
\newcommand{\asca}{{\it ASCA~}}
\newcommand{\rosat}{{\it ROSAT~}}
\newcommand{\rxj}{RXJ1248.4+0831~}

\makeatletter

\newenvironment{inlinefigure}{%
\def\@captype{inlinefigure}%
\noindent\begin{minipage}{\linewidth}\begin{center}}
{\end{center}\end{minipage}\smallskip}
\makeatother

\shorttitle{\chandra Observations of NGC4698}
\shortauthors{Georgantopoulos \& Zezas}

\begin{document}
\title{\chandra observations of NGC4698: a Seyfert-2 with 
 no absorption}
   
\author{I.~Georgantopoulos$\!$\altaffilmark{1} 
A.~Zezas$\!$\altaffilmark{2}  
}

\altaffiltext{1}{Institute of Astronomy \& Astrophysics, 
National Observatory of Athens, Palaia Penteli, 15236, Athens, Greece}
\altaffiltext{2}{Harvard-Smithsonian Center for Astrophysics, 60 
 Garden Street, Cambridge, MA02138}

\begin{abstract}
 We present \chandra ACIS-S observations of the enigmatic 
 Seyfert-2 galaxy NGC4698. This object together 
 with several other bona-fide Seyfert-2 galaxies show no absorption 
 in the low spatial resolution \asca data, in contrast to the standard 
 unification models. Our \chandra observations of NGC4698 probe directly 
 the nucleus allowing us to check whether nearby sources contaminate 
 the \asca spectrum.  
 Indeed, the \chandra observations show that the 
 \asca spectrum  is dominated by two nearby AGN. 
 The X-ray flux of NGC4698 is dominated by 
 a nuclear source with luminosity $\rm L_{0.3-8 keV}\sim10^{39}$ \lunits , 
 coincident with the radio nucleus.
 Its spectrum is well represented by a power-law, $\Gamma\approx 2.2$, 
 obscured by a small column density of $5\times10^{20}$ \cunits
 suggesting that NGC4698 is an atypical Seyfert  galaxy. 
 On the basis of its low luminosity we then interpret NGC4698 as a Seyfert  
 galaxy which lacks a broad-line region. 
\end{abstract}



\section{Introduction}

  Seyfert galaxies are roughly divided in those 
 presenting broad emission lines with $>1000 \rm km~s^{-1}$,
 (Seyfert-1)  
 in their spectra and those having only narrow lines (Seyfert-2). 
 In the standard Active Galactic Nuclei (AGN),
 unification models  (see Antonucci 1993 for a review)
 the two types of Seyfert galaxies are intrinsically  identical with 
 their differences being an  orientation effect. 
 In particular, an accretion disk produces the 
 UV continuum which ionizes the surrounding gas:
 the Broad Line Region (BLR) and the Narrow Line Region (NLR)  
 at distances typically $<$0.1 pc and $<$ 100 pc respectively
 (e.g. Robson 1996).  
 Then a Seyfert galaxy is classified as type-2  
 if our line of sight toward the nucleus 
 intercepts an obscuring screen,
 the so-called torus, which blocks the BLR.  
 There have been several observations 
 supporting these simple unification models. 
 Ground breaking optical studies have detected 
 broad lines in polarized light in NGC1068 
 demonstrating  
 that a hidden BLR is present 
 in this Seyfert-2 galaxy (Antonucci \& Miller 1985). 
 Similarly IR observations showed the existence of obscured 
 BLRs in several Seyfert-2 galaxies (Veilleux et al. 1997).  
 Further evidence supporting the unification 
 models comes from X-ray observations of Seyfert-2 galaxies which  
 show large amounts of obscuration,  
 typically above $10^{23}$ \cunits, 
 e.g. Smith  \& Done (1996), Turner et al. (1997). 
 
 However, some questions arise 
 regarding whether the unification models can be applied to all 
 Seyferts.  Polarimetry studies have shown that 
 a significant fraction of Seyfert-2 galaxies  
 appear to lack a hidden BLR (Moran et al. 2000, Tran 2001). 
 A correlation with luminosity is observed in the 
 sense that the less luminous Seyfert nuclei are those 
 missing a BLR (Tran 2001, Lumsden \& Alexander 2001). 
 Moreover, \asca observations of some  
 Seyfert-2 galaxies show that these do not  present 
 intrinsic X-ray absorption: NGC3147, Ptak et al. (1996),
 NGC7590, Bassani et al. (1999), NGC4698, Pappa et al. (2001). 
 Recently Panessa \& Bassani (2002) added several more  
 candidates in this list.  One has to be cautious 
 in identifying candidate 'unabsorbed Seyfert-2' as 
 the optical spectroscopy may  
 not be of sufficient quality in order to classify them as 
 bona-fide Seyfert-2 galaxies. 
 Furthermore, some of the objects may be  Compton thick i.e. the 
 transmitted component below 10 keV is
 completely absorbed. Then in the \asca 
 1-10 keV band we would observe only the 
 scattered, {\it unabsorbed} component. 
 The lack of a strong Fe line at 6.4 keV as well as the
 high ratio of the broad band hard X-ray  over the [OIII] line  
 emission (the latter comes from the NLR and thus 
 represents an isotropic measurement of the luminosity; e.g. 
 Alonso-Herrero et al. 1997) 
 can be used to discriminate against the Compton thick model. 
 Finally, the \asca Point Spread Function is large 
 (3 arcmin Half-Power Diameter) and hence 
 the observed X-ray emission may be contaminated by other nearby sources, or the integrated emission of 
the host galaxy.  
 This confusion problem is 
 particularly important in the Low Luminosity 
 Seyferts where the energy output of a few strong Ultra-Luminous 
 X-ray (ULX) sources may rival the luminosity of the nucleus 
 (e.g. Ho et al. 2001).

 Here, we discuss \chandra observations of NGC4698.
 NGC4698, at a distance of 16.8 Mpc,  belongs to the 
 spectroscopic sample of nearby galaxies 
 of Ho, Filippenko \& Sargent (1997). 
 The excellent quality nuclear spectra 
 obtained ($2\times4$ arcsec slit)
 make the Seyfert-2 classification 
 for this galaxy highly certain.
 Pappa et al. (2001) observed 
 NGC4698 with \asca. Their best-fit 
 model is an unobscured 
 ($9\pm 4 \times 10^{20}$ \cunits)
 power-law $\Gamma=1.91^{+0.12}_{-0.10}$
 with a luminosity of $\rm L_{2-10 keV}\approx 2\times 10^{40}$ \lunits. 
 No FeK line is detected with the 
 upper limit on equivalent width being
 $\sim 0.4$ keV at the 90\% confidence level. The absence of a strong 
 FeK line as well as the high value of the 
 $f_x/f_{[OIII]}$ ratio suggested that NGC4698 is 
 not a Compton thick object. 
 However, the spectrum measured by \asca 
 may be contaminated by other nearby sources 
 which mask the true nuclear emission.  
 The excellent 
 spatial resolution of \chandra allows us to 
 obtain a direct view of the nuclear region without any   
 confusion problems.

\section{Observations and Data Reduction}
 NGC4698 was observed on 2002-June-16 (OBSID 3008),  using 
 the Advanced CCD Imaging Spectrometer, ACIS-S,
 onboard {\it Chandra} (Weisskopf et al. 1996). 
 The back-illuminated ACIS-S3 CCD was on the aimpoint, providing
 a spatial resolution of  $0.5"$ and minimizing Charge Transfer 
 Inefficiency (CTI) problems.
 In order to minimize pile-up the  observation was performed in  
 $1/2$-subarray mode.
 In this mode the field of view is $4.1'\times8.3'$ which 
 covers the whole galaxy.  
 We use the type-2 event file (including only grades 0,2,3,4 and 6) provided 
 by the standard pipeline processing  
 after discarding periods of high background. 
 The resulting   exposure time is 29726 sec. 
 Images, spectra an lightcurves have been created  
 using the {\sl CIAO v2.2} software.  
 The imaging and timing analysis  
 were performed using the {\sl SHERPA} 
 software. For the spectral fitting 
 we use the {\sl XSPEC v11} software package.

\section{Analysis}

\subsection{Imaging}
 Three strong sources are detected by \chandra in the vicinity of 
 NGC4698. The detected sources together with the 
 \asca GIS contours  are shown in Fig.1. 
 The cross gives the position 
 of the central source which is spatially coincident 
 with the optical center of the galaxy 
 and therefore we identify it as the nucleus.
 The nuclear source is also coincident with a faint radio source 
 detected by Ho \& Ulvestad (2001) at  
 equatorial coordinates $\rm \alpha=12^h48^m22^s.92$  
 $\rm \delta=+08^\circ 29^m14''.5$ (J2000). 
 Its  radio flux density was  0.23 mJy at 6 cm corresponding to a radio power 
 of $\sim 10^{19}$ $\rm W~Hz^{-1}$.  
 \rxj (source 1,  on Fig. 1), originally detected in the \rosat All-Sky-Survey 
 was identified as a redshift z=0.12  AGN (Bade et al. 1998).
  J124825.9+083021  is a radio source (denoted as source 2 on Fig.1 ) 
 detected by Ho \& Ulvestad (2001), having 
 a flux density of 0.8 mJy at 6 cm.
 Recently, Foschini et al. (2002) identified this source
 as a probable BL Lac at a redshift of z=0.43.   
 Since the latter sources are $\sim6$ and $\sim4$ times 
 brighter than the nucleus of NGC4698
 in the 0.3-8.0~keV band (Table~1), 
  the \asca spectrum is dominated 
 by them {\it instead of the NGC4698 nucleus}. 
 Although, the \asca data cannot 
 give an uncontaminated spectrum of the nucleus 
 due to their limited spatial resolution,  
 Pappa et al. (2001) assumed that all the X-ray emission observed by ASCA arises from
 NGC~4698 and failed to 
 discuss the possible contamination from the nearby 
 \rosat source.   

 The X-ray luminosity of the nuclear source is relatively low 
 ($\rm L_{X}\sim 10^{39}$ \lunits; 0.3-8.0~keV), leaving open the possibility 
 that it could be associated with a ULX,
 (see Ward 2002 for a recent review). 
 Nevertheless, the fact that the nuclear X-ray source is coincident with the 
 radio and optical nucleus suggests that our 
 source is most probably associated with the supermassive black 
 hole at the center of NGC4698. 
  Moreover, the ratio of the X-ray to radio luminosity (see Terashima \& Wilson 
 2003 for the definition) of the 
 nuclear source in NGC4698 is
 an order of magnitude lower than in the case of the ULX in NGC5408, the only 
 extragalactic X-ray binary emitting at
 ULX levels identified with a radio source (Kaaret et al. 2003).
 The value of the above ratio ($\sim 10^{-6}$) puts NGC4698 clearly in the 
 regime of radio quiet AGN.   

 Seven more X-ray sources are detected within or around NGC4698
 down to $\sim2 \times 10^{-15}$ \funits.
 The details of all the above sources are given in Table 1. 
 We give the name, equatorial coordinates (J2000), counts, 
 the absorbed flux and luminosity (all in the total 0.3-8 keV band) in
 columns 1, 2, 3, 4 and 5 respectively.
 Counts were extracted from a 4 pixel (2 arcsec) radius circle.  
  For the seven off-nuclear sources in NGC4698, the conversion 
 from count rate to flux was performed assuming 
 a power-law spectrum of $\Gamma=1.9$, consistent with the spectrum 
 of Low Mass X-ray Binaries 
 in nearby galaxies e.g. Prestwich et al. (2003).  
 We further assume that the  above spectrum is only 
 absorbed by the Galactic column 
 density, $\rm N_H=2\times10^{20}$\cunits, (Dickey \& Lockman 1990).  
 For the nucleus, \rxj and  J124825.9+083021   we used the spectra 
 derived from the spectral fits below. 
 Luminosities are estimated assuming $\rm H_o=75~km~s^{-1}~Mpc^{-1}$,
 $\rm q_o=0.5$ and $\rm \Lambda=0$.  

 We have fitted the two-dimensional spatial profile of the 
 nuclear source with the {\sl SHERPA} software. 
 We use a radius of 2 arcsec for our fit. 
 Fitting a Gaussian profile, we obtain a FWHM of $1.58\pm 0.18$ pixels.
 Given that the nominal Point Spread Function (PSF) is undersampled by the 
 pixels of the ACIS camera (FWHM$_{PSF}\sim0.5''$ at 1.5~keV, 
 compared to a pixel size of  $0.49''$) 
 we consider the probability of any extension very marginal.

\subsection{Spectral Fitting} 
 We obtained spectra for the three brightest sources, for 
 which is possible to perform any spectral analysis, using an 
extraction radius of 2 arcsec.
 Background regions are taken from source free regions 
 on the same chip. Note however, that the background  contribution is 
 practically negligible with about 1 count in the above
 extraction radius.   
 We group the data so that there are at least 15 counts per bin. 
 The quoted errors to the best fitting spectral parameters 
 correspond to the 90\% confidence level for one interesting 
 parameter.
 We discard data below 0.3 keV 
 due to the low response as well as calibration uncertainties.
  In order to take into account the degradation of the ACIS quantum 
 efficiency in low energies, we used the {\sl ACISABS} model in the 
 spectral fitting\footnote{http://asc.harvard.edu/cal/Acis/Cal\_prods/qeDeg}.
 We present spectral fits for the following sources:

\underline{The nucleus of NGC4698} 
 A single power-law fit gives a good fit 
 to the data ($\chi_\nu^2=5.8/9$). The column density is 
 $\rm N_H=5^{+0.7}_{-0.5}\times 10^{20}$\cunits~ a factor 
 of two above the Galactic value 
 ($2.0\times10^{20}$\cunits; Dickey \& Lockman, 1990).   
 This is much lower than the column densities 
 encountered in typical Seyfert-2 galaxies 
 (eg Smith \& Done 1996, Turner et al. 1997).
 The photon index is  $\Gamma=2.18^{+0.28}_{-0.44}$.    
 Unfortunately our data are not of sufficient quality 
 to check for the presence of an FeK line at 6.4 keV. 
 The spectrum of the source together with the 
 residuals from the best-fit power law model are 
 given in Fig. 2.

\underline{\rxj}
The single power-law fit yields a relatively poor fit to the data 
($\chi^2=75.8/55$). The photon index is $\Gamma\sim 1.7$
 while the column density is constrained to be  less than 
 $1.0\times 10^{20}$\cunits.
 The photon index is characteristic of those 
 of Seyfert-1 nuclei (Nandra \& Pounds 1994).  
 When we add a thermal Raymond-Smith component 
 (Raymond \& Smith 1977),  absorbed only by the Galactic column, 
 the fit is significantly improved ($\chi^2=64.5/55$).
 The temperature of the thermal component is $\rm kT\sim 0.2$ keV.  
 Then, the photon index becomes $\Gamma=1.49^{+0.12}_{-0.12}$. 
 The luminosity of the thermal component is 
 $\rm L_{0.3-2 keV} \sim 6\times 10^{41}$ \lunits or about 20 per cent of the 
 total luminosity in this band. 

\underline{J124825.9+083021}  
 A single power-law fit yields $\chi^2_\nu = 40.0/38$. 
 The column density is consistent with the Galactic
 ($\rm N_H < 8\times 10^{20}$ \cunits) while the photon 
 index is $2.02\pm 0.11$.    
 The \chandra spectrum in very good agreement 
 with that obtained by Foschini et al. (2002).
These authors find $\Gamma=2.0\pm 0.2$ with $\chi^2_\nu = 18.3/24$
 using {\it XMM-Newton} data. Note that the X-ray flux has increased 
 by less than a  factor of two between the \chandra and the 
 {\it XMM-Newton} observation (obtained at 16/12/2001).

\section{Discussion}
 Previous 'large-beam'  observations,  
 mostly performed with \asca and {\it BeppoSAX},
  have shown the presence of 
 a population of low luminosity Seyfert-2 galaxies 
 with no absorption in their X-ray spectra
 (e.g. Panessa \& Bassani 2002).
 However, the low X-ray spatial resolution  
 casts doubt on whether the X-ray spectra observed 
 are those of the nuclei. 
 The problem is more acute in the case of 
 low luminosity nuclei where the 
 nuclear flux can be easily outshined by adjacent 
 bright sources.      
 Indeed, the \chandra observation of NGC4698 
 reveals  that the \asca data 
 {\it did not probe the nucleus of NGC4698}. 
 The X-ray flux in the \asca beam is   
 dominated by two nearby sources
 while the contribution of the NGC4698 nucleus is very small, $\sim10$ 
 per cent of the total 0.3-8 keV \asca flux. 
 We find that several other sources in the galaxy contribute about 
 40 per cent of the total galactic flux. 
 Given their luminosity 
 ($\rm L_x\sim 5\times10^{37}-2\times10^{38}$ \lunits), these 
 are most probably associated with X-ray binaries. 
 However, accidentally, the spectrum of the nucleus 
 again appears to show no absorption 
 having $\rm N_H=5\times 10^{20}$ \cunits, only a factor of 
 two above the Galactic column density. 
 However, as the nuclear flux measured by 
 \asca is erroneous, we need to assess again whether 
 NGC4698 is a Compton thick source. 
 Using the new \chandra 2-10 keV flux, 
 the $\rm f_{2-10 keV}/f_{[OIII]}$ ratio is $\sim1$. 
 This value is still in the regime of the 
 Compton thin AGN (Bassani et al. 1999). 
 Although  the FeK$\alpha$ line 
 could provide additional information, 
 the poor photon statistics of NGC4698 do not 
 allow us to set any constraints based on this diagnostic tool:
 no photons have been detected above 5 keV. 
 Nevertheless, we note that Compton thick Seyferts
 appear to have a flat spectrum $\Gamma\sim 1 $ 
 in the 2-10 keV band (e.g. Matt et al. 1996).
 This can be explained as the hard reflection component 
 from the back-side of the torus  dominates the observed 
 spectrum in the 2-10 keV band over the 
 reflected power-law spectrum. 
 A pure reflected $\Gamma=1.9$ spectrum,
 in principle, could 
 be obtained in the rare case where the orientation 
 is such that the torus 
 is exactly edge-on so that the hard reflection 
 component is completely hidden from view.
 Therefore, the \chandra spectrum rather 
 adds evidence against the Compton thick scenario.  

 To our knowledge, NGC4698 is the second Seyfert-2 galaxy 
 from the Ho et al. (1997) sample with no 
 X-ray absorption as demonstrated by \chandra observations.  
 The other example  is NGC3147: 
 a recent snapshot \chandra observation of this galaxy 
 (Terashima \& Wilson 2002), 
 clearly shows that the nucleus is unabsorbed 
 in agreement with the \asca spectrum. 
 In particular, the  
 \chandra spectrum yields $\Gamma=1.79^{+0.17}_{-0.09}$
 with $\rm N_H\sim1.5\times10^{21}$ \cunits
 ($\rm L_{2-10 keV}\sim 8\times 10^{41}$ \lunits). 
 No FeK line is detected at 6.4 keV with \chandra.  
 Moreover, the variability observed between the \chandra 
 and the \asca epoch strongly suggests that the X-ray emission 
 is not scattered radiation from an extended region. 
  
 A  model where  the X-rays 
 come through  unabsorbed while 
 the BLR is attenuated is rather unlikely. 
 A warm absorber
 retaining dust could in principle attenuate 
 the optical radiation while leaving intact the 
 X-rays. Still, a warm absorber would imprint strong 
 oxygen absorption features in the X-ray spectrum below 1 keV.
 Such absorption edges have not been seen in the 
 spectrum of NGC3147, while it is impossible to 
 check for such features in the much poorer quality  
 spectrum of NGC4698.  
 Another possible explanation for the lack of 
 X-ray absorption in NGC4698 and NGC3147 is that we are 
 simply viewing  a Seyfert-1 nucleus 
 which does not have a BLR. This is the pure Seyfert-2 model
 of Tran (1995).  
 Recent models (eg Nicastro 2000, Laor 2003) 
 suggest that a BLR cannot exist 
 at low accretion rates. In particular, 
 Nicastro (2000) (see also Nicastro, Martocchia \& Matt 2003)
 advocate a model where the BLR is associated with a wind 
 coming from the accretion disk. In this model the 
 width of the broad lines depends on the accretion rate
 in the sense that AGNs accreting in a 
 {\it much lower rate than their Eddington limit } 
 ($< 1-4\times 10^{-3}\times~M_{Edd}$)
 should possess no BLR.
 Note however, that the exact accretion rate 
 limit depends on the accretion disk physics.
 Therefore it may  not be surprising 
 that there are a few cases of low accretion rate AGN
 with a BLR (see the discussion in Laor 2003). 
 
 The accretion rate
 in the case of  NGC4698 can be easily estimated as follows. 
 The observed velocity dispersion  is 
 169 $\rm km~s^{-1}$ (McElroy 1995; Corsini et al. 1999, estimate a somewhat lower value 
 of 134 $\rm km~s^{-1}$). 
 Then  from the relation between the 
 black hole mass and the velocity dispersion 
 (e.g. Gebhardt et al. 2000),  
 we find a black-hole mass of $\rm M_{BH}\sim 2-8\times10^7 ~ M_\odot$. In this estimate we
 take into account both the quoted errors in the 
 relation of Gebhardt et al. (2000) as well 
 as the variations between the different measurements of the  velocity dispersion. 
 Hence, the predicted Eddington luminosity is 
 $\sim 3-10 \times 10^{45}$ \lunits.
 We estimate the bolometric luminosity by 
 integrating the measured X-ray luminosity  down to 100$\mu m$ 
 using a power-law  energy spectrum of slope $\alpha=1$, characteristic 
 of the broad-band AGN spectrum (e.g. Robson 1996). This yields 
 $\rm L_{BOL}= 4\times10^{39}$ \lunits.      
 Note however, that if we use instead the 
 relation between the bolometric luminosity 
 and the narrow $H\alpha$ luminosity (see Laor 2003) 
 we obtain $\rm L_{BOL}=4\times10^{40}$ \lunits.
 Hence $\rm L_{BOL}/L_{EDD}\sim 4\times 10^{-7}-1 \times 10^{-5}$. 
 In the case of NGC3147 we estimate a central mass of 
 $3-5\times 10^8 ~M_\odot$, 
 from  the measured bulge velocity dispersion, 
 268 $\rm km~s^{-1}$, (McElroy 1995). 
 Therefore the predicted Eddington Luminosity 
 is $\rm L_{EDD}=4-6\times 10^{46}$ \lunits.  
 Extrapolation of the X-ray emission down to 100$\mu m$, using 
 $\alpha=1$ gives  $\rm L_{BOL}=3\times 10^{42}$ \lunits; 
 the narrow $H\alpha$ luminosity gives instead 
  $\rm L_{BOL}=3\times 10^{41}$ \lunits. Hence, 
 $\rm L_{BOL}/L_{EDD}\sim 5\times10^{-6}-1\times10^{-4}$.     
According to  the  theoretical models discussed earlier, these low accretion
 rates are totally consistent with an absent BLR. 
 
 Unfortunately, there are no polarimetry data 
 for NGC4698 or NGC3147 in order to search for the presence of a hidden BLR
 and therefore to test directly the above model.
 In the case of NGC7590, which also has a low column density,  
 where such data are available 
 (Heisler, Lumsden \& Bailey 1997) no BLR is detected. 
 Tran et al. (2001) find that the 
 Seyferts with no hidden BLR are those with
 the lowest luminosity.  Therefore,  
 it is possible that there is a link between  
 the Seyfert-2 galaxies which present no 
 absorption in their X-ray spectrum 
 with those which lack a hidden BLR.   
 However, in the samples of Tran et al. (2001) and 
 Lumsden \& Alexander (2001) there are 
 a few examples of Seyfert-2 galaxies without a BLR but with 
 large amounts of X-ray absorption. 
 This could be simply explained according to 
 an orientation effect in the standard 
 unification model framework.   
 Assuming that all AGN with low accretion rate  
 do possess a torus but have no BLR, 
 objects like NGC4698 which present no X-ray absorption 
 must be viewed face-on
 while the ones with absorption are viewed edge-on. 

 In conclusion,  \chandra observations of NGC4698 
 (and NGC3147) confirm that these galaxies
 host nuclei which present very little X-ray absorption, although 
 they are classified as Seyfert-2 by optical spectroscopy.
 This is in disagreement with the standard unification models 
 where a dense obscuring screen is believed to block the BLR 
 and to absorb the soft X-ray radiation. 
 The most straightforward explanation is that NGC4698 
 lacks a BLR. This could be in line with theoretical models 
 where the BLR clouds are not formed at low accretion rates, 
 and explains the low absorbing column density in other 
 low-luminosity AGNs classified as 
 type-2 on the basis of their optical spectra.

\acknowledgements
We thank Fabrizio Nicastro for many useful comments. 
This work has been supported by the NASA grant
NAG-G02-3127X.



\begin{center}
\begin{table}[h]
\scriptsize
\caption{The detected X-ray sources 
\label{tab}}
\begin{tabular}{ccccc}
\tableline\tableline
Name    &  Equatorial Coordinates & Counts & logf (0.3-8 keV) & logL (0.3-8 keV) \\
        &       (J2000)           &        &          \funits & \lunits\\
\hline 
Nucleus & 12 48 22.9 +08 29 15 & 168           & -13.5     & 39.0 \\
CXUJ124823.6+082844     & 12 48 23.6 +08 28 44 &28  & -14.3 & 38.2 \\
CXUJ124823.3+082901     & 12 48 23.3 +08 29 01 &10  & -14.8 & 37.7 \\
CXUJ124823.2+082915     & 12 48 23.2 +08 29 15 & 14 & -14.6 & 37.9 \\
CXUJ124822.2+082917     & 12 48 22.2 +08 29 17 & 24 & -14.4 & 38.1 \\
CXUJ124822.2+082926     & 12 48 22.2 +08 29 26 & 14 & -14.6 & 37.9 \\
CXUJ124824.3+082754     & 12 48 24.3 +08 27 54 & 12 & -14.7& 37.8 \\
CXUJ124826.4+082955     & 12 48 26.4 +08 29 55 & 14 & -14.6 & 37.9 \\
        &                      &    &     &      \\
RXJ1248.4+0831  & 12 48 28.4 +08 31 12 & 986&  -12.6& 42.8    \\
 J124825.9+083021 & 12 48 25.9 +08 30 21 & 699&  -12.8& 43.8    \\
\hline   
\tableline
\end{tabular}
\end{table}
\end{center}

\vspace{3.0cm} 

\begin{center}
\begin{table}[h]
\scriptsize
\caption{Power-law Spectral fits  
\label{tab}}
\begin{tabular}{cccc}
\tableline\tableline
Source    &  $\rm N_H$ ($10^{20}$\cunits) & $\Gamma$ & $\chi^2$ \\ \hline 
NGC4698 nucleus &  $5^{+0.7}_{-0.5}$ & $2.18^{+0.28}_{-0.44}$ & 5.8/9 \\
\rxj      & $<1$ & $1.67^{+0.09}_{-0.09}$ & 75.8/55 \\
J124825.9+083021 & $4^{+4}_{-4}$ & $2.02^{+0.11}_{-0.11}$ & 40.0/38 \\ 
\hline 

\hline 
\tableline
\end{tabular}
\end{table}
\end{center}

\eject 

\begin{inlinefigure}
\epsscale{0.5}
\plotone{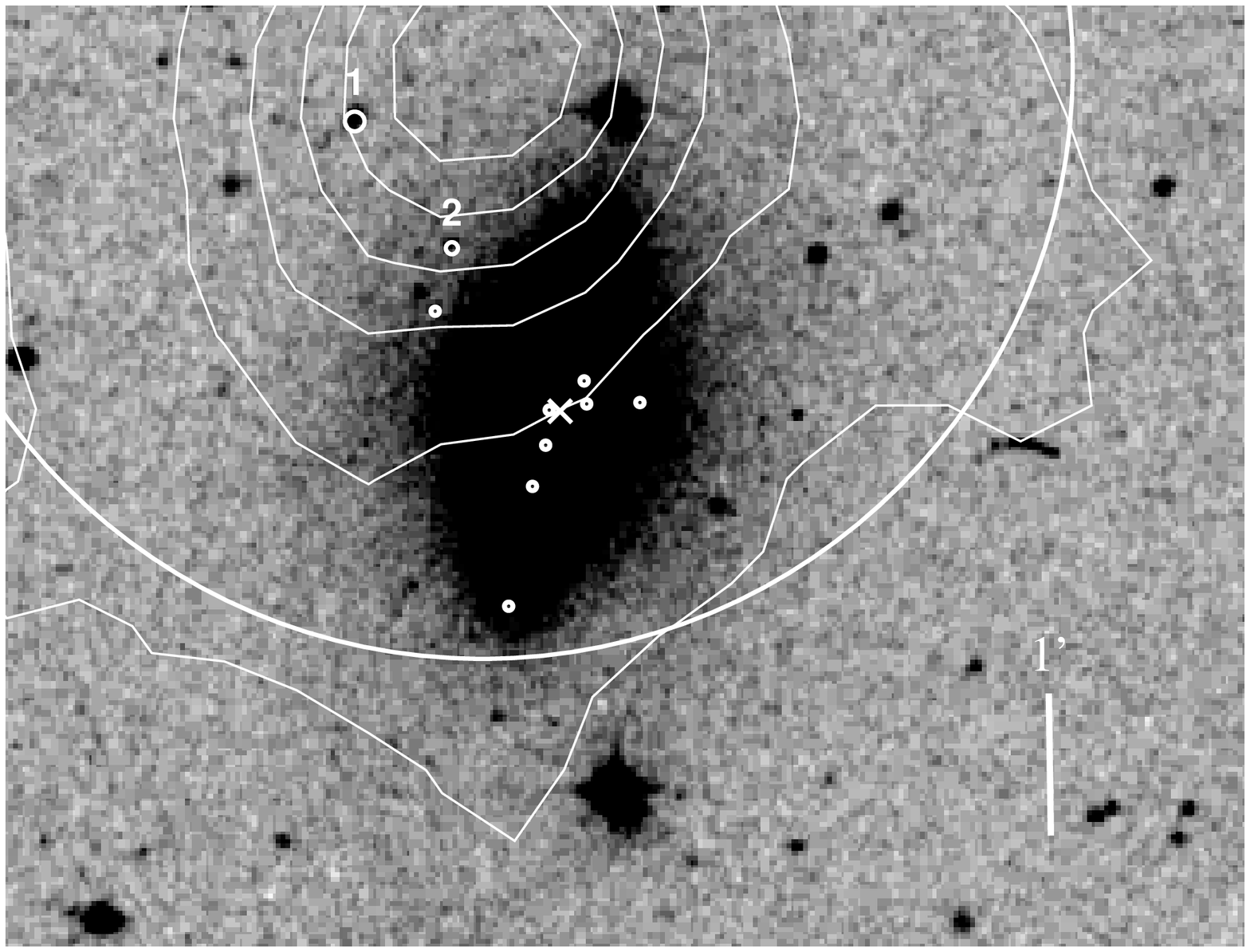}
\plotone{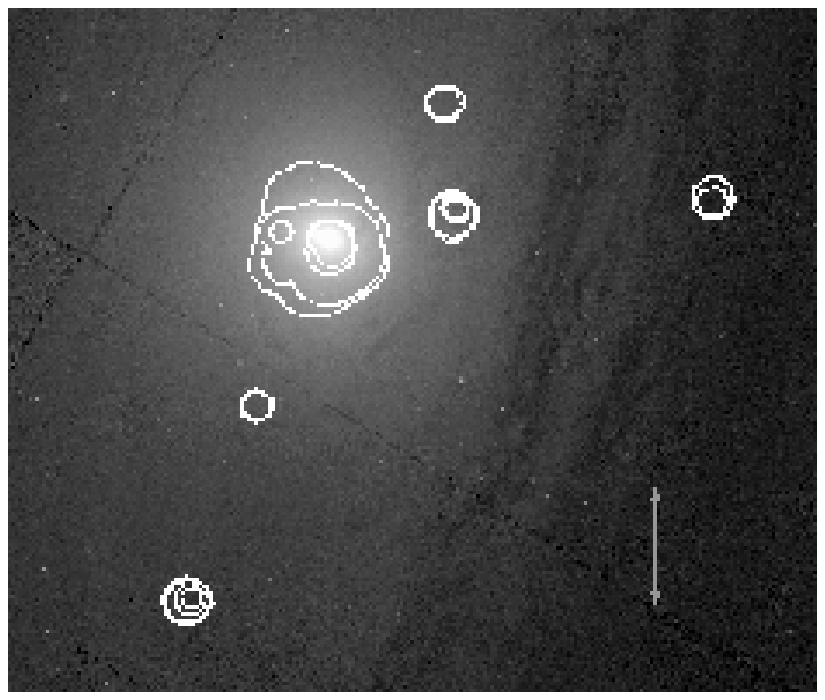}
\figcaption{Upper: The sources detected in the \chandra image overlaid 
 on the Digital Sky Survey image. Sources 1 and 2 denote 
 \rxj and the radio source J124825.9+08302 respectively. 
 The white dots correspond to the 7 sources which are 
 probably associated with the galaxy, given in table 1. 
 The cross denotes the most luminous source in NGC4698 which 
 coincides with the optical and the radio nucleus.
 The contours from the \asca GIS observation are also overlaid.   
 The circle (4 arcmin radius) 
 corresponds to the \asca extraction radius used by 
 Pappa et al. (2001).   
 The line corresponds to 1 arcmin. 
 Lower: The \chandra 0.3-8 keV contours overlaid
 on an {\it HST} WFPC2 (F606W filter) image. The line corresponds to 
 10 arcsec.} 
\label{image}
\end{inlinefigure}

\begin{inlinefigure}
\label{image}
\epsscale{0.5}
\rotatebox{270}{\plotone{f2.eps}}
\figcaption{The spectrum of the nuclear source in 
 NGC4698 together with the best fit power-law model 
 (upper panel). In the lower panel we plot 
 the residuals of the data from the best fit model.
}
\end{inlinefigure}

\end{document}